\documentclass[review]{elsarticle}

\usepackage{amssymb}
\usepackage{subfigure}
\journal{Astroparticle Physics}

\begin{document}

\begin{frontmatter}

\title{A New Mirror Alignment System for the VERITAS Telescopes}

\author{A.~McCann\corref{cor1}}
\author{D.~Hanna}
\author{J.~Kildea}
\author{M.~McCutcheon}

\cortext[cor1]{Corresponding author: mccann@physics.mcgill.ca}
\address{Physics Department, McGill University, Montreal, QC H3A 2T8, Canada}

\begin{abstract}
Imaging atmospheric Cherenkov telescopes (IACTs) used for ground-based
gamma-ray astronomy at TeV energies use reflectors with areas on the
order of 100m$^2$ as their primary optic.  These tessellated
reflectors comprise hundreds of mirror facets mounted on a space frame
to achieve this large area at a reasonable cost.  To achieve a
reflecting surface of sufficient quality one must precisely orient
each facet using a procedure known as alignment.  We describe here an
alignment system which uses a digital (CCD) camera placed at the focus
of the optical system, facing the reflector.  The camera acquires a
series of images of the reflector while the telescope scans a grid of
points centred on the direction of a bright star.  Correctly aligned
facets are brightest when the telescope is pointed directly at the
star, while mis-aligned facets are brightest when the angle between
the star and the telescope pointing direction is twice the
misalignment angle of the facet.  Data from this scan can be used to
calculate the adjustments required to align each facet.  We have
constructed such a system and have tested it on three of the VERITAS
IACTs.  Using this system the optical point spread functions of the
telescopes have been narrowed by more than 30\%.  We present here a
description of the system and results from initial use.

\end{abstract}

\begin{keyword}
VERITAS, IACT, Alignment, Optics
\end{keyword}

\end{frontmatter}
\section{Introduction}
The current generation of imaging atmospheric Cherenkov telescopes
operating around the world~\cite{Holder08,Hinton04,Baixeras04} has
ushered in a new era in TeV gamma-ray astronomy.  The number of
detected TeV gamma-ray sources has grown from below ten in 2000 to
more than seventy today~\cite{Aharonian08} largely because of the
increased sensitivity of the instrumentation.  This increase results
from the use of the following:
\begin{itemize}
\item{larger reflectors}
\item{cameras with larger fields of view and higher resolution}
\item{multiple telescopes making stereoscopic observations}
\item{flash-ADC-based data acquisition systems}
\end{itemize}
 
For the benefits afforded by large reflectors and high resolution
cameras to be fully realised, the optical quality of the telescopes
must be maintained at a high level.  Since the reflectors of these
telescopes comprise several hundred mirror facets, their alignment
presents a significant technical and logistical challenge.

The VERITAS array, located in southern Arizona, USA, employs four
twelve-metre-diameter f-1.0 reflectors of the Davies-Cotton
type~\cite{DaviesCotton}.  Each reflector consists of a tubular steel
optical support structure (OSS) on which 345 identical hexagonal
mirror facets are mounted.  The facet mounts allow precision
adjustments to bring the focus of each to the same point on the
primary focal plane of the telescope.  Since the mirror facets are
exposed to the dust of the Arizona Sonoran desert their reflectivity
degrades over time so they are therefore re-coated on a regular
basis~\cite{Roache08}.  This process maintains the reflectivity of the
facets but their removal and re-installation compromises the optical
quality of the reflector as a whole so the alignment of the facets
must be repeated on a regular basis.

We present here an alignment system, based on the technique originally
suggested by Arqueros et al.~\cite{Arqueros05}, which can be used for
aligning the VERITAS telescopes.  It achieves the quality desired in a
reasonable length of time at modest cost.  Importantly, the optimal
alignment is achieved for typical observation elevations.

\section{Method and Apparatus}
\begin{figure}
\centering
\subfigure[On axis]
{
\includegraphics[width=0.47\textwidth]{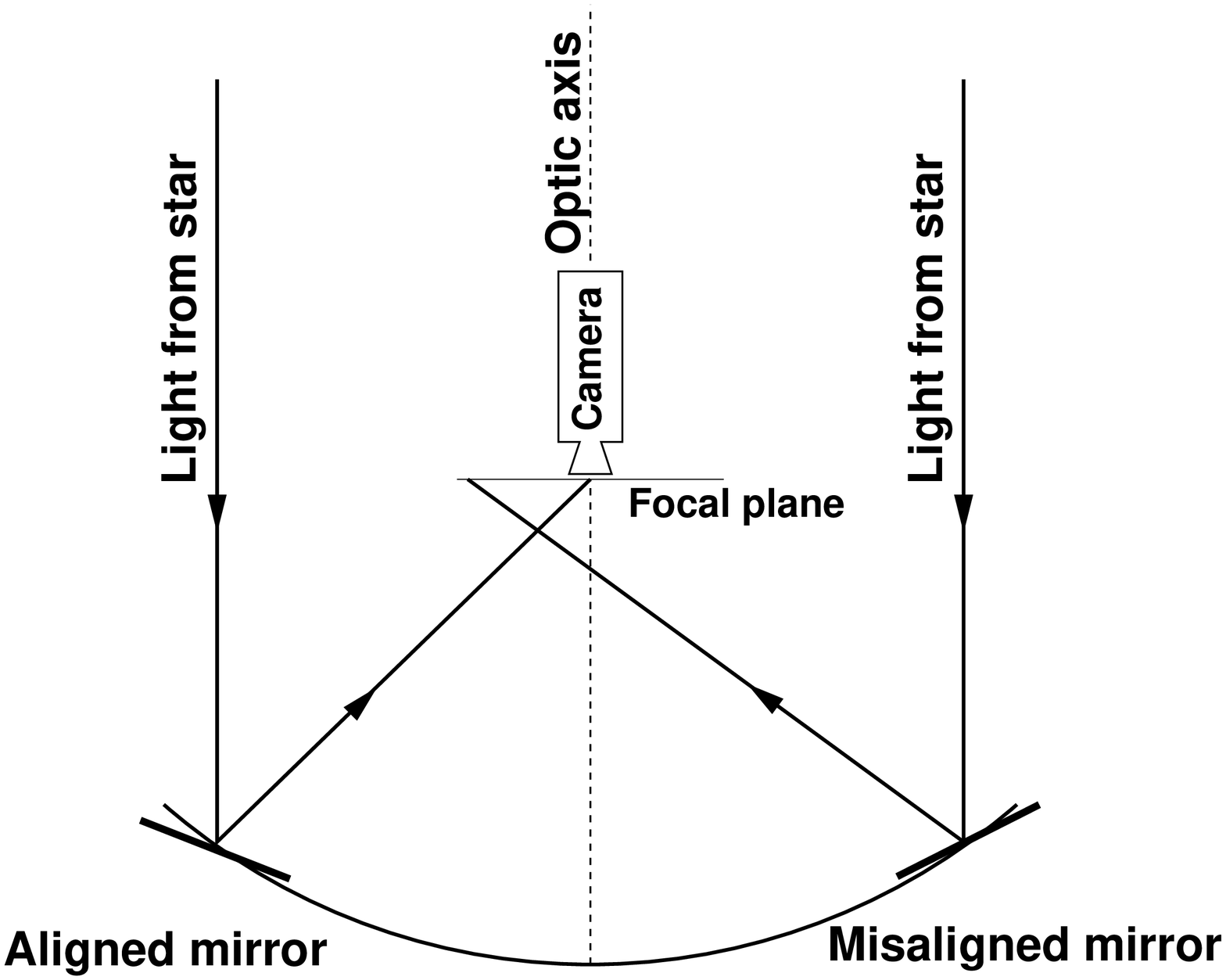}
}
\subfigure[Off axis]
{
\includegraphics[width=0.47\textwidth]{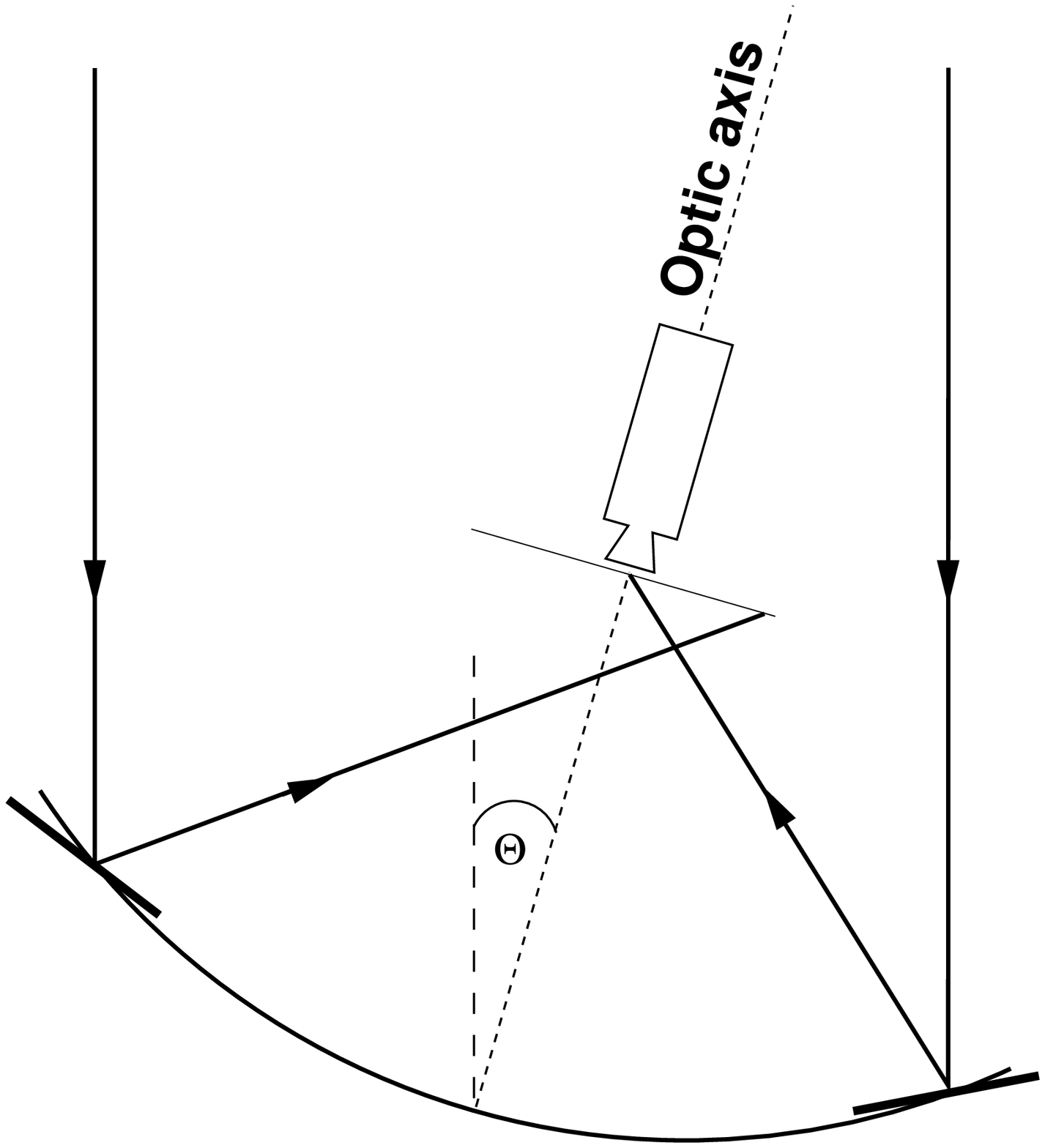}
}
\caption{ An illustration of the raster scan method. In panel (a) the
  well-aligned facet will appear bright in the CCD-camera image while
  the mis-aligned facet will be dark.  In panel (b) the mis-aligned
  facet will appear brightest when the angle between the star and the
  telescope pointing direction, $\theta$, is twice the misalignment
  angle of the facet.  }
\label{fig:cartoon}
\end{figure}

\begin{figure}[]
  \centering \includegraphics[width=0.7\textwidth]{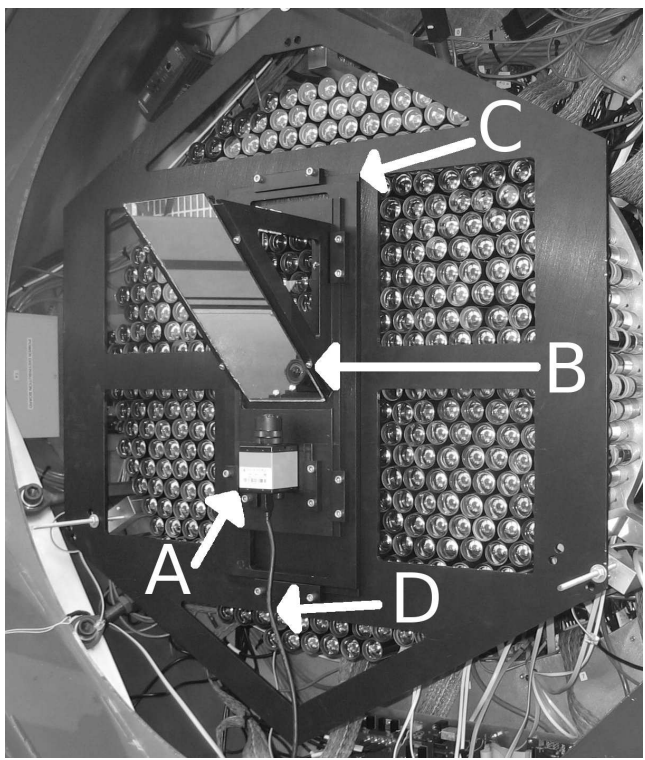}
  \caption{ A photograph of the alignment apparatus mounted on one of
    the VERITAS PMT cameras.  Arrow A indicates the digital camera; B,
    the 45$^{\circ}$ mirror; C, the x-y positional stage and D the
    cable connecting to the data acquisition notebook computer (not
    shown).  }
  \label{fig:app}
\end{figure}

Our alignment system uses a digital camera which is mounted at the
centre of the telescope's focal plane, facing the reflector.  Images
of the reflector are acquired at each point of a raster scan that the
telescope performs centred on a bright star at a typical observing
elevation.  At each point in the raster scan, the camera registers the
amount of light from each facet; the point in the scan at which a
given facet appears brightest occurs when the angle between the
pointing direction of the telescope and the star is exactly twice the
mis-alignment angle of the facet (see Figure~\ref{fig:cartoon}).  On
completion of the raster scan the acquired images are analysed and
correction adjustments are calculated for each facet.
A photograph of the alignment apparatus is shown in Figure~\ref{fig:app}.
The apparatus consists of:
\begin{itemize}
\item{a mounting plate}
\item{an x-y positional stage}
\item{a 45$^{\circ}$ plane mirror}
\item{a digital camera with wide-angle lens}
\item{a notebook computer}
\end{itemize}

The mounting plate is constructed from 6-mm anodised aluminum and has
several large cut-outs to reduce its weight.  It enables quick and
reproducible installation on any of the VERITAS telescopes with no
modifications to the photomultiplier tube (PMT) camera required.  The
camera and plane mirror are mounted on the x-y positional stage which
allows the camera's virtual image in the 45$^{\circ}$ mirror to be
located on the telescope's optical axis at the prime focus of the
reflector.  The camera, model DMK 21BF04 from Imaging Source, is based
on a 1/4-inch, 640$\times$480 pixel, monochrome CCD device.  The
wide-angle f-1.4 lens is a Computar T2314-FICS-3 with a 2.3 mm focal
length and a 22.8 mm effective front aperture.

\begin{figure}[]
\centering \subfigure[] {
  \includegraphics[width=0.47\textwidth]{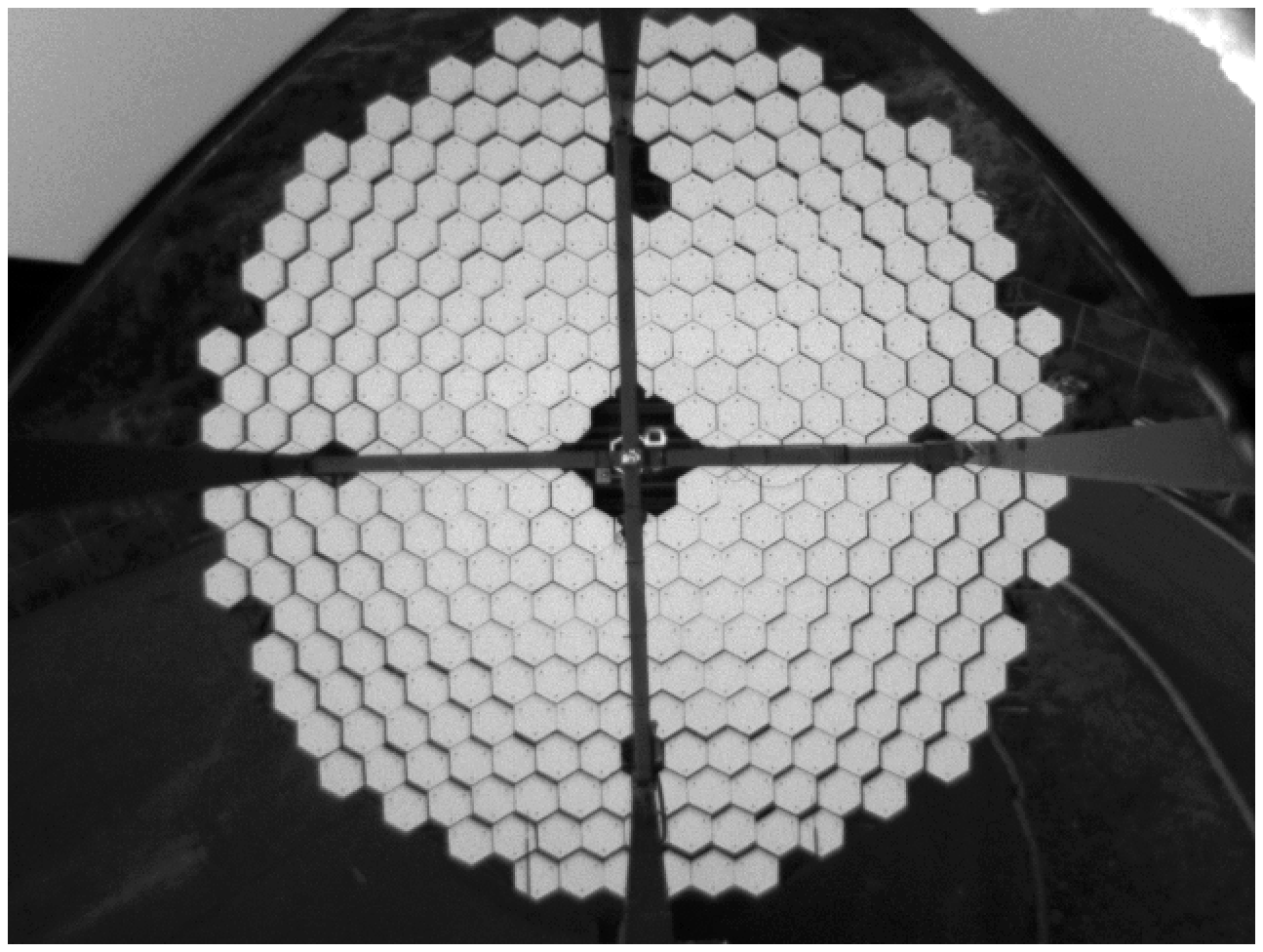} }
\subfigure[] {
  \includegraphics[width=0.47\textwidth]{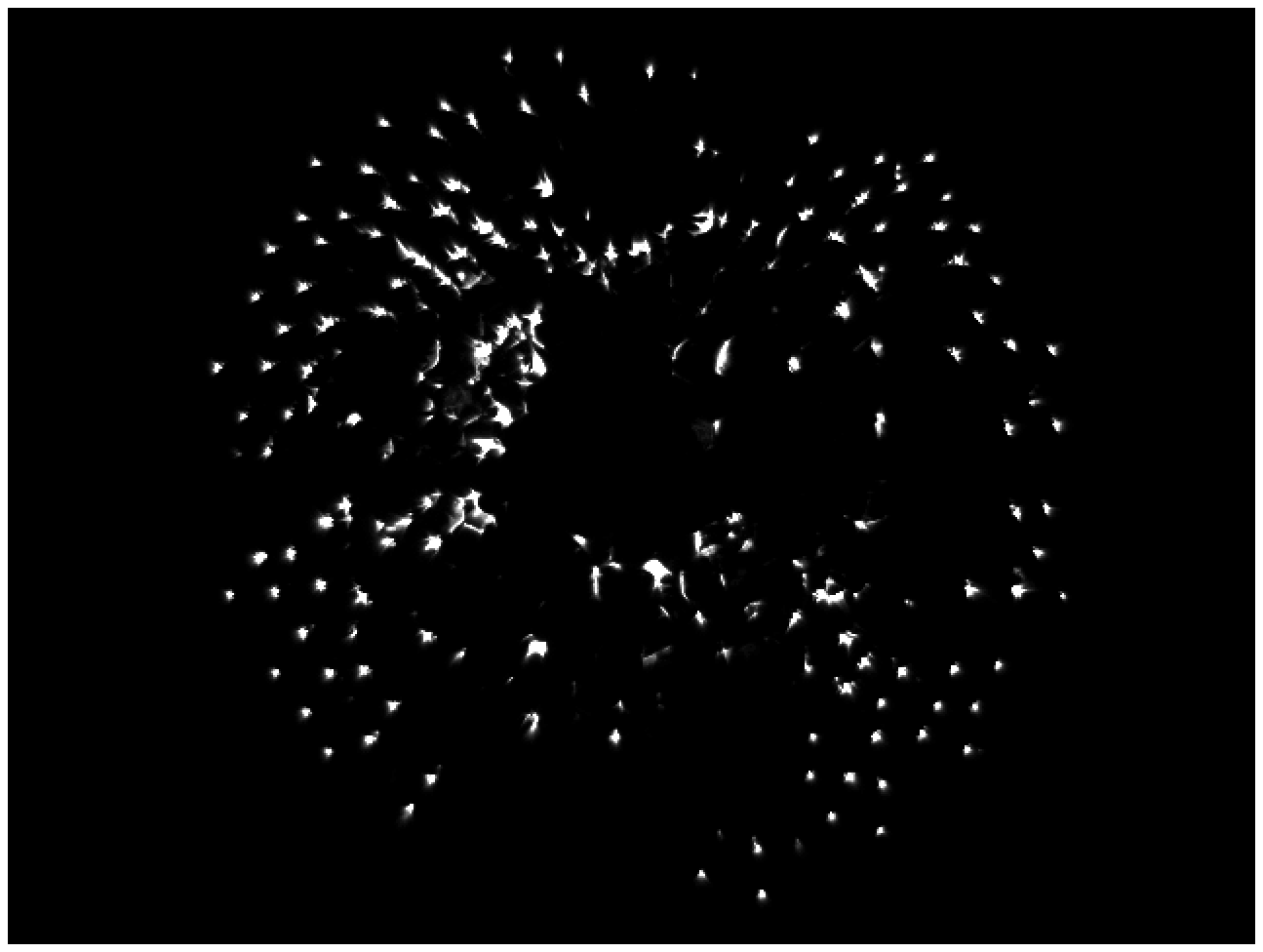} }
\caption{
Images, taken by the digital camera, of one of the VERITAS reflectors during 
the day (a) and at night while pointing at a star (b). 
The bright spots in the right-hand panel are caused by starlight reflecting off 
well-aligned facets while the dark regions indicate poorly-aligned facets.
}
\label{fig:dish}
\end{figure}

Image acquisition software runs on a notebook computer which is
connected to the camera via a firewire interface and to the telescope
tracking computer via ethernet.  Images are stored in the FITS data
format~\cite{Hanisch01} with the telescope pointing information saved
in the image metadata.  Two images of the telescope reflector, taken
with the alignment camera, are displayed in Figure~\ref{fig:dish}.

\section{Data acquisition and analysis}
The first stage of the data acquisition process is the recording of a
template image: an image of the reflector fully illuminated with all
facets clearly visible (see Figure~\ref{fig:dish}a).  This image is
used to map between the pixels of the CCD camera and the facets of the
reflector.  More precisely, a circular region inside each facet,
encompassing $\sim$90\% of the CCD pixels associated with the facet,
is selected.  In the analysis of all subsequent images the signal in
these pixels is assumed to be caused by light reflecting off the
corresponding facet.  The signal in the remaining $\sim$10\% of the
facet is ignored in the analysis.  This region may be illuminated by
light reflecting off the edge of the facet or may be contaminated by
the signal from pixels illuminated by the neighbouring facet bleeding
across the CCD.  A template must be recorded every time the alignment
system is mounted or adjusted in order to ensure that the mapping
between the CCD pixels and the reflector facets is accurate.  An image
of a VERITAS reflector, taken at twilight, is used as the template
image for data acquired during the following night.  In cases where
the raster-scan data are acquired after a partial night of standard
gamma-ray observations, the template image is acquired with the moon
illuminating the reflector.

The second stage of the data acquisition consists of capturing
successive images of the reflector while the telescope performs a
raster scan centred on a star of magnitude 3 or brighter.  In the
tests presented here we used stars which transited at an elevation of
$\sim$70$^{\circ}$ and tracked them for two hours, with tracking
beginning one hour before culmination.  This ensured that the entire
scan was performed over a small elevation range ($<$ 5$^{\circ}$).
This is necessary because the OSS deforms slightly under gravity as
the telescope moves in elevation so categorisation and optimisation of
the telescope's optics should be performed over a range of elevations
close to those used for most astronomical observations.

The raster scans used for the tests reported on here were performed
over a grid of 21$\times$21 pointings, on the plane tangent to the
right ascension and declination of the chosen star.  The angular
spacing between each row and column in the grid was 0.02$^{\circ}$.  A
program running on the telescope's tracking computer was used to slew
the telescope to the required grid coordinate.  A pause of three
seconds was then observed, to allow any post-slewing oscillations of
the telescope to die out, before the CCD camera was commanded to
capture an image of the reflector.  Once the capture process was
completed the telescope was slewed to the next grid point. This
combination of grid size, grid resolution and telescope settling time
was chosen to allow a scan which could be completed in two hours and
which scanned an area fully encompassing the point-spread-function
(PSF).

\begin{figure}[]
\centering \subfigure[] {
  \includegraphics[width=0.47\textwidth]{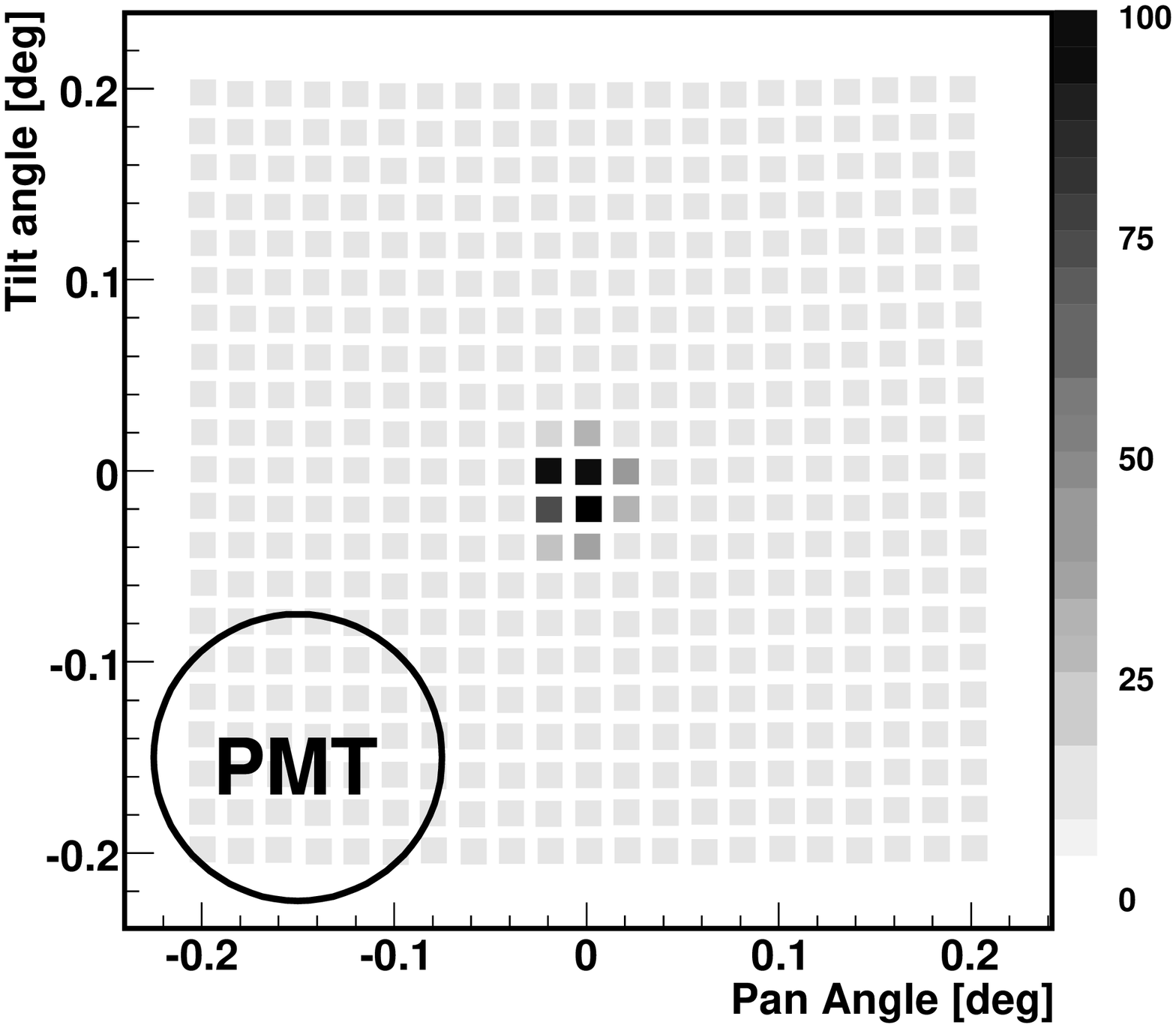} } \subfigure[] {
  \includegraphics[width=0.47\textwidth]{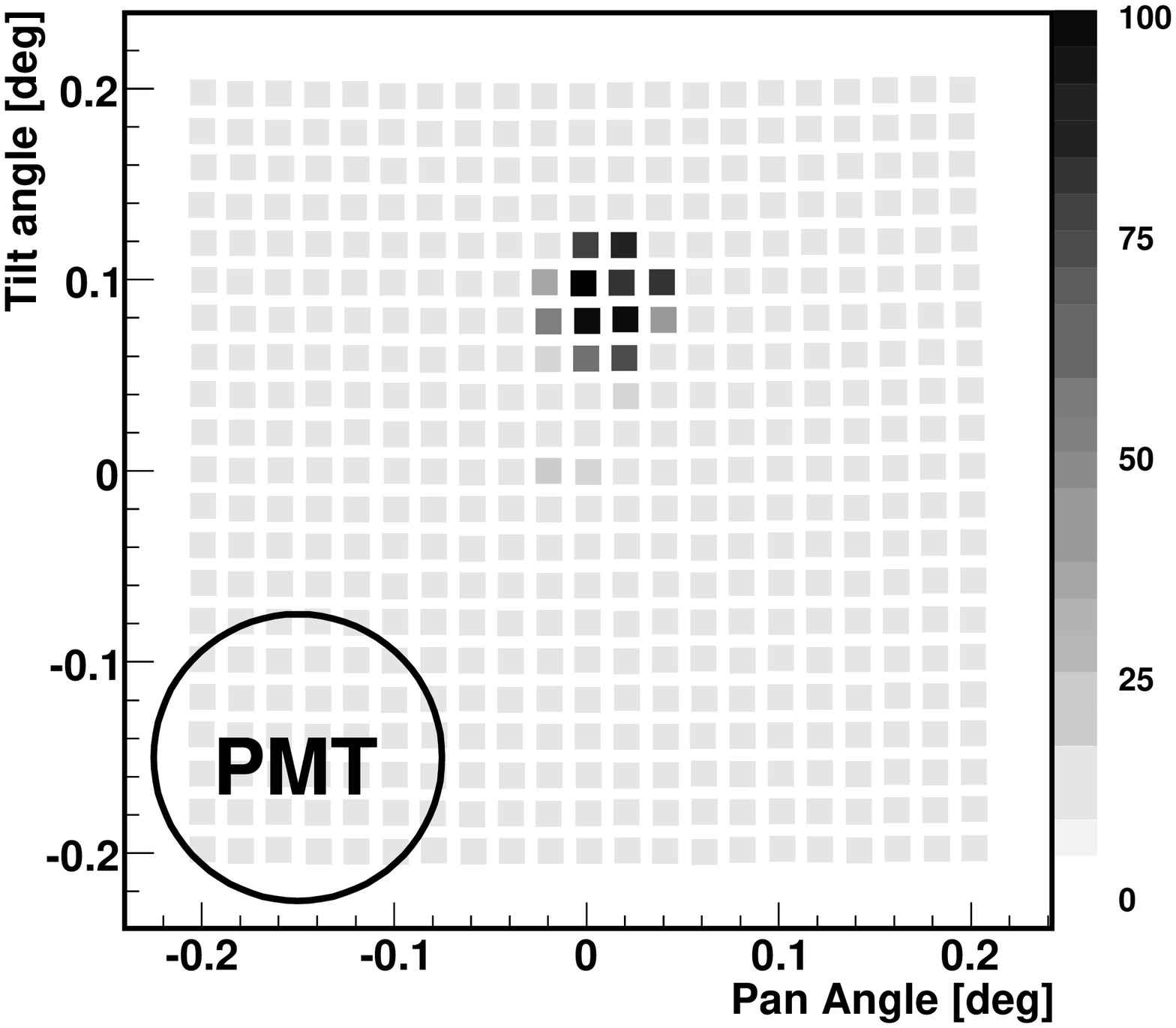} }
\caption{ Plots of facet brightness at different telescope pointing
  offsets for two facets.  The left panel corresponds to a facet which
  is most effectively illuminated close to the centre of the raster
  grid, where the telescope is pointing directly at the target star.
  This indicates a well-aligned facet.  The right panel corresponds to
  a facet which is mis-aligned since it exhibits its best illumination
  when the angle between the telescope pointing direction and the star
  is $\sim$ 0.1$^{\circ}$.  This facet must be tilted downwards by
  $\sim$ 0.05$^{\circ}$ to be correctly aligned.  The brightness value
  (greyscale) is plotted in arbitrary units.  The black circle
  indicates the size of a PMT in the VERITAS camera.  }
\label{fig:grid}
\end{figure}

The images of the reflector, captured during the scan, are analysed in
the following way.  For each image, a brightness value is assigned to
every facet. The brightness value associated with a given facet is
calculated by summing the signal in the pixels which correspond to the
facet, as determined from the map generated from the template image.
These brightness values are then plotted at the corresponding scan
points in a two-dimensional map (see Figure~\ref{fig:grid} for
examples).  The scan point with the maximal brightness for a given
facet identifies the mis-alignment angle of the facet.  The telescope
pointing offset for that scan point corresponds to twice the
mis-alignment angle since the angle of incidence and angle of
reflection of the starlight change together.

In practice, the positioning of the alignment apparatus on the
telescope optical axis is not perfect.  This leads to a small
systematic bias in the calculated mis-alignment angles which manifests
itself as a non-zero value for the the mean of all mis-alignment
angles.  This bias is subtracted from the mis-alignment angles when
computing the alignment adjustment values.

\begin{figure}[t]
\includegraphics[width=\textwidth]{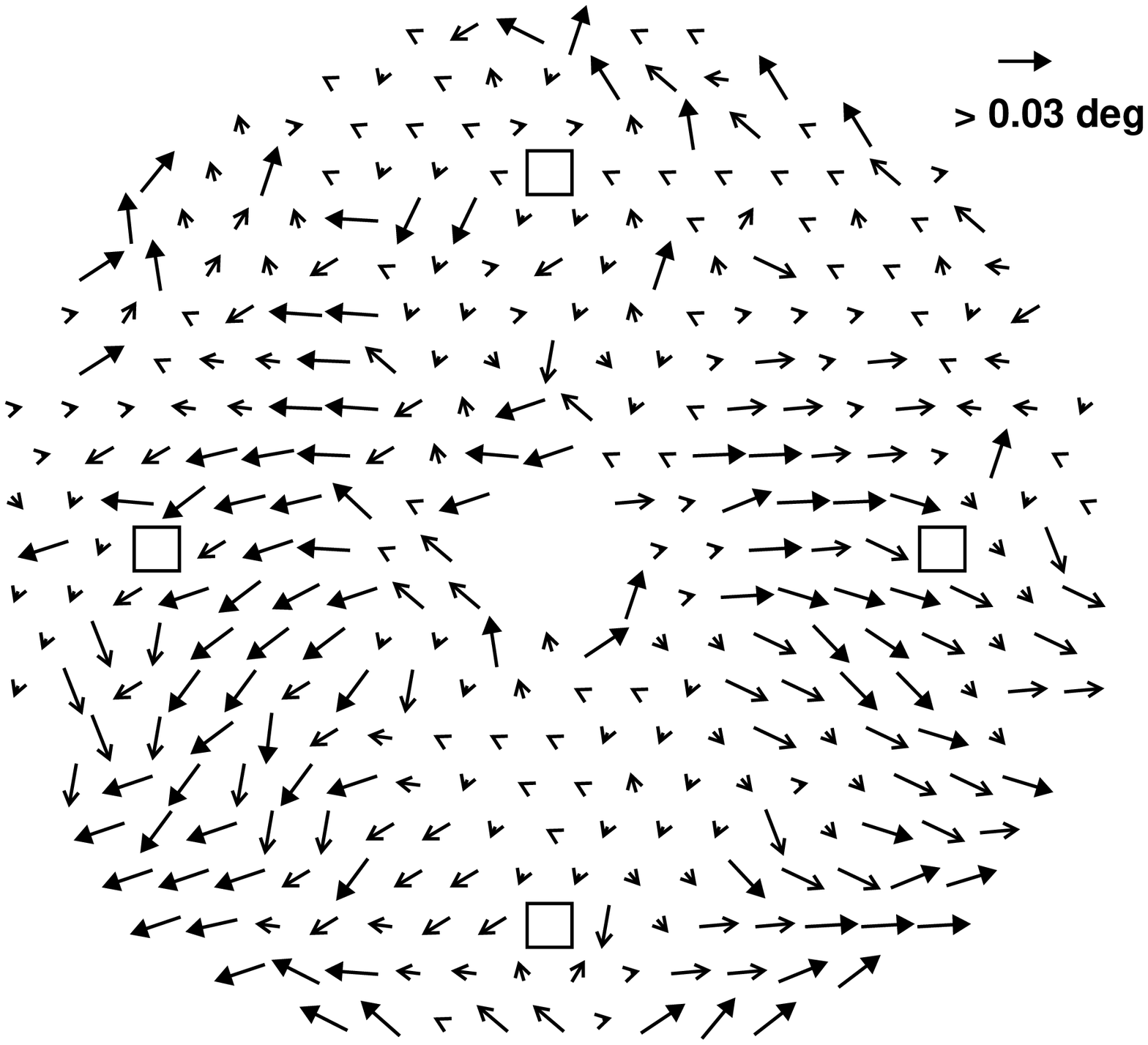}
\caption{ A mis-alignment map for one of the VERITAS telescopes.  The
  length of the arrow indicates the size of the misalignment of an
  individual facet.  Misalignments angles greater than 0.03$^\circ$
  are not drawn to scale and are plotted with a solid arrow head.  The
  black squares indicate the position of the quadrupod arms of the
  telescope.  }
\label{fig:arrows}
\end{figure}

The mis-alignments determined from a raster scan can be summarised in a single
plot, as shown in Figure~\ref{fig:arrows}.

\section{Correction Implementation}
Each mirror facet of a VERITAS reflector is supported by a triangular
three-point suspension.  At each vertex, a brass mounting gimbal and
adjustment nut are threaded onto a stainless-steel threaded rod.  Any
mis-alignment of the facet can be corrected by turning two of these
adjustment nuts.  The mirror-mount geometry and threaded-rod pitch are
such that one full turn on a nut changes the mirror orientation by
$\sim$0.1$^{\circ}$. The adjustments computed from the raster scan
data were implemented on the VERITAS mirrors manually with a
socket-wrench device which had a circular index wheel attached to it.
This allowed adjustments as small as 1/16th of a turn
($\sim$0.007$^{\circ}$) to be accurately implemented.

The adjustments were implemented during daylight hours following the
raster scan procedure.  Experience shows that tuning up an already
nominally aligned telescope following the replacement of 50 facets
takes only a few hours.
 
\section{Alignment results}
The optical quality of the VERITAS telescopes has improved with the
implementation of this alignment system.  Three of the four telescopes
were aligned during May, 2009 (the fourth was being dismantled for
relocation at that time) and the size of the PSF was reduced by more
than 30\% from previous values.  The 80\% containment radii of the
PSFs are now less than $\sim$0.05$^{\circ}$ at operational elevations.

\begin{figure}[]
\centering
\subfigure[PSF image before]
{
\includegraphics[width=0.47\textwidth]{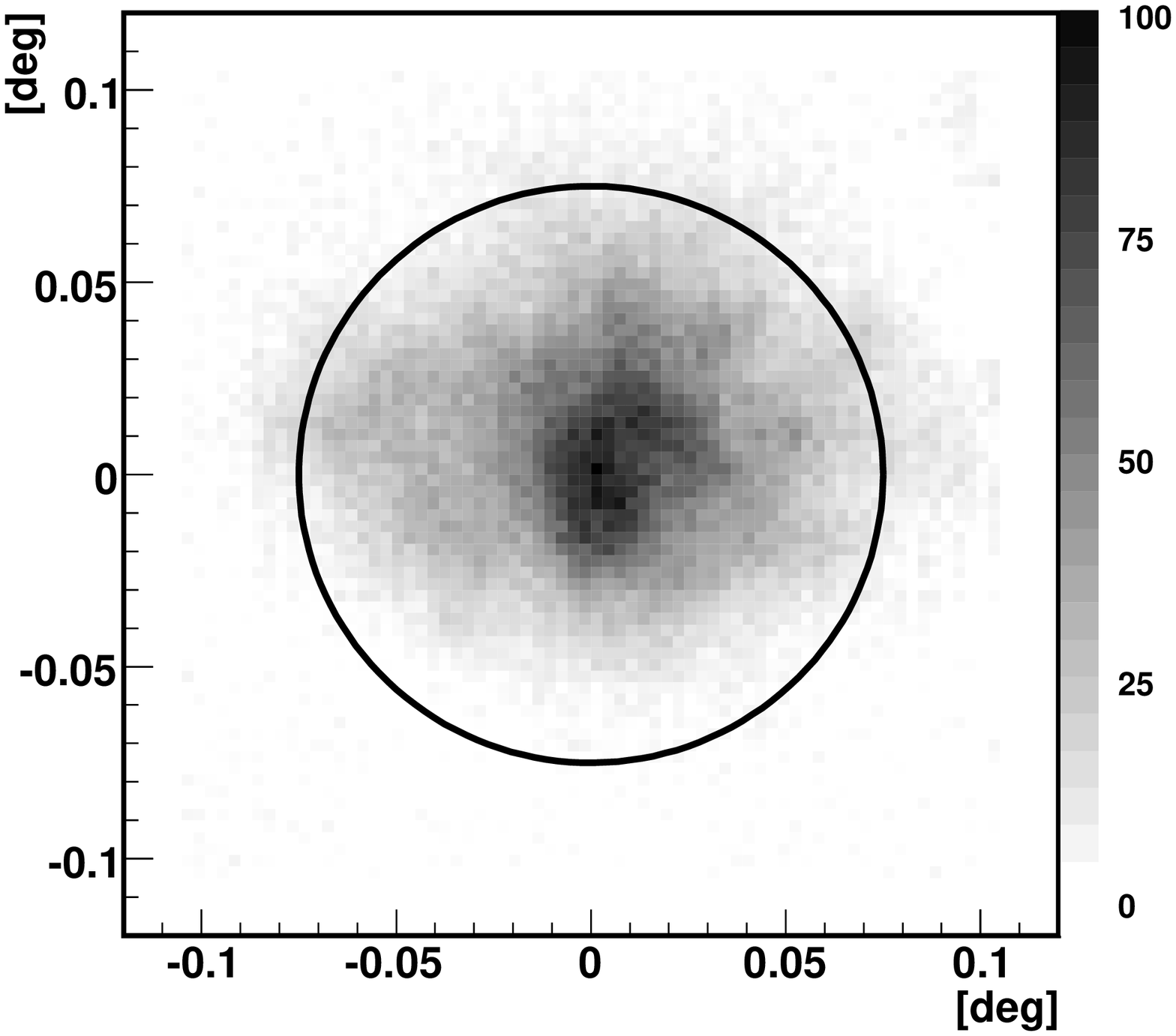}
}
\subfigure[PSF image after]
{
\includegraphics[width=0.47\textwidth]{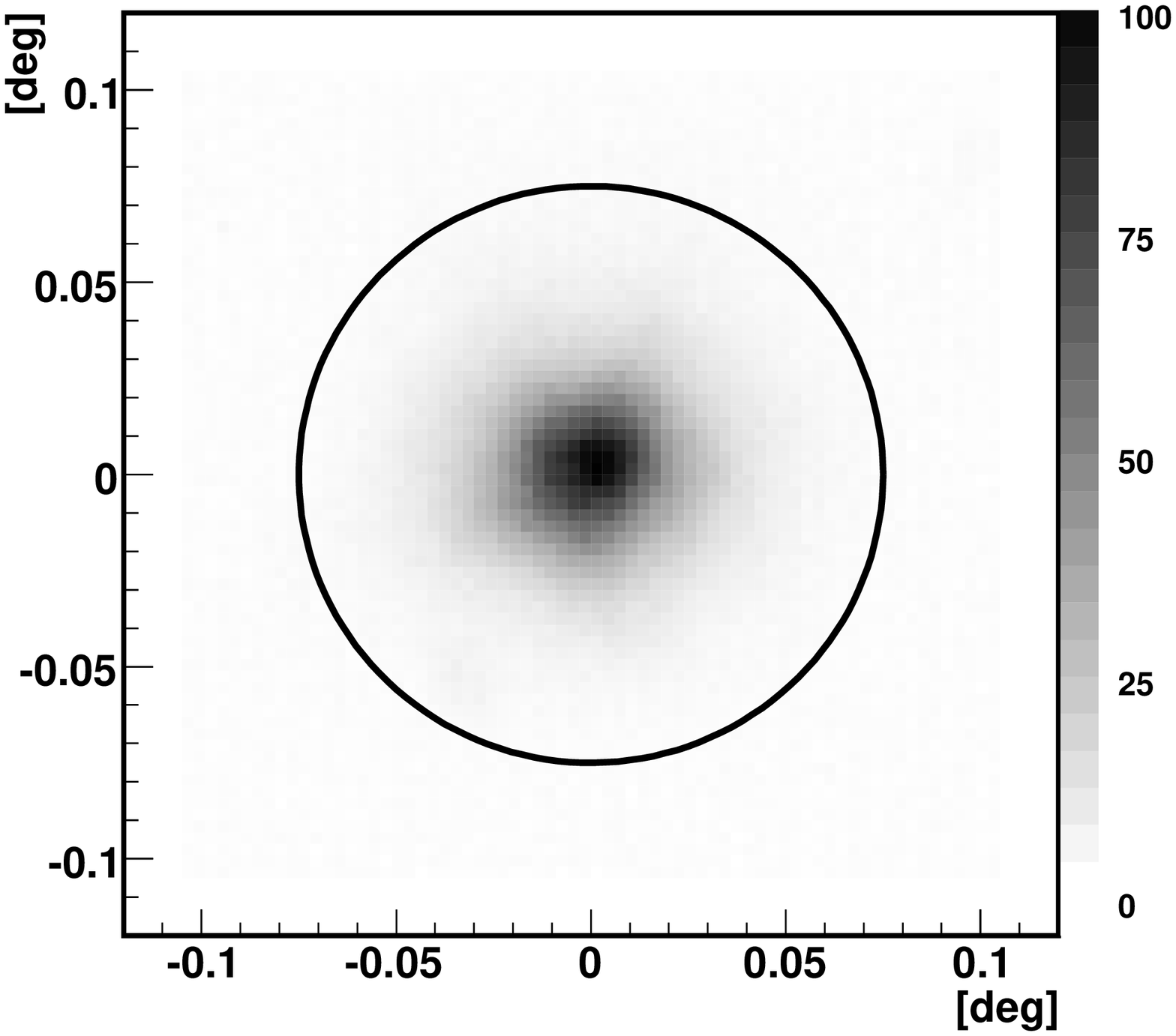}
}
\subfigure[Reflector image before]
{
\includegraphics[width=0.47\textwidth]{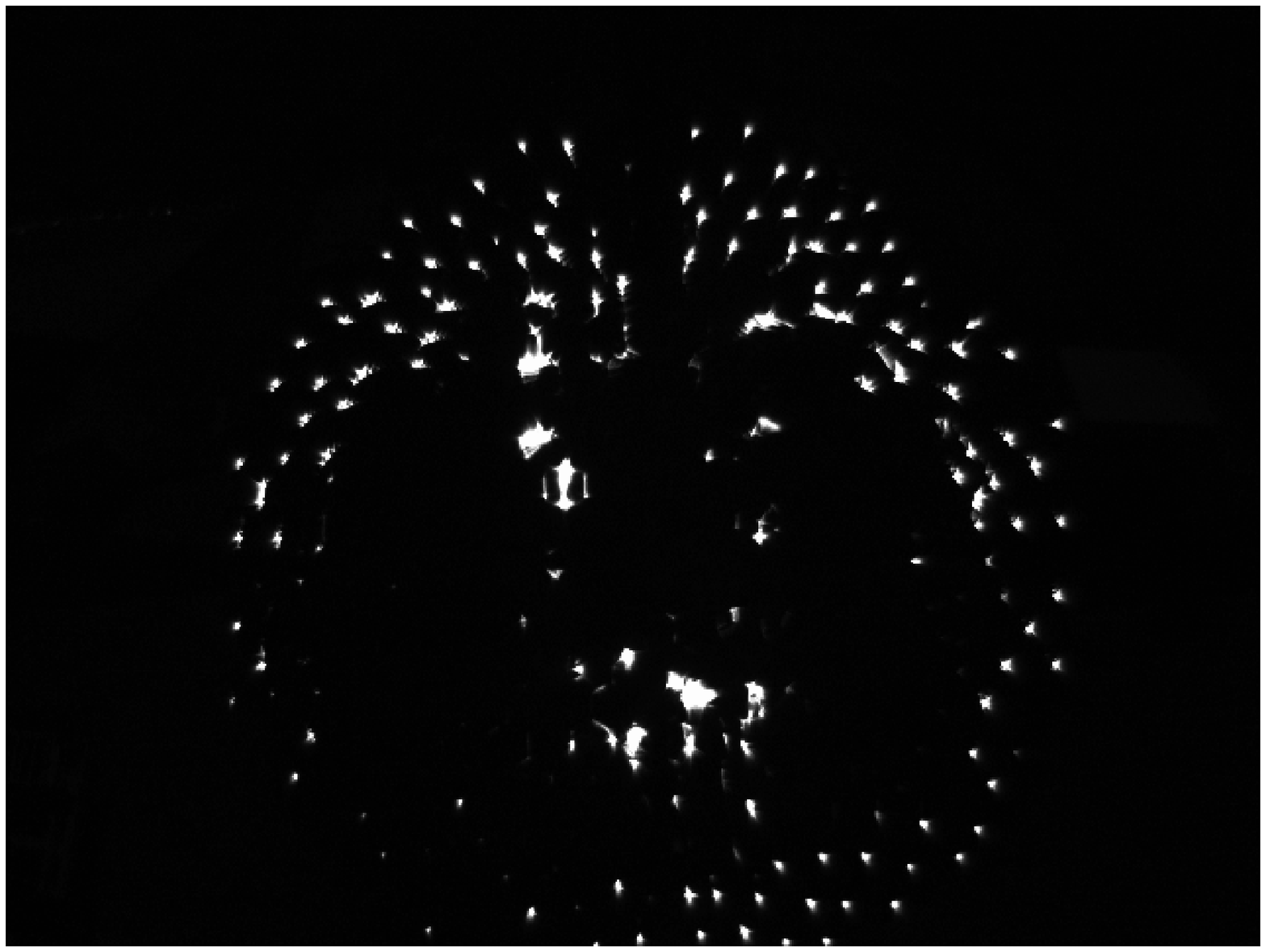}
}
\subfigure[Reflector image after]
{
\includegraphics[width=0.47\textwidth]{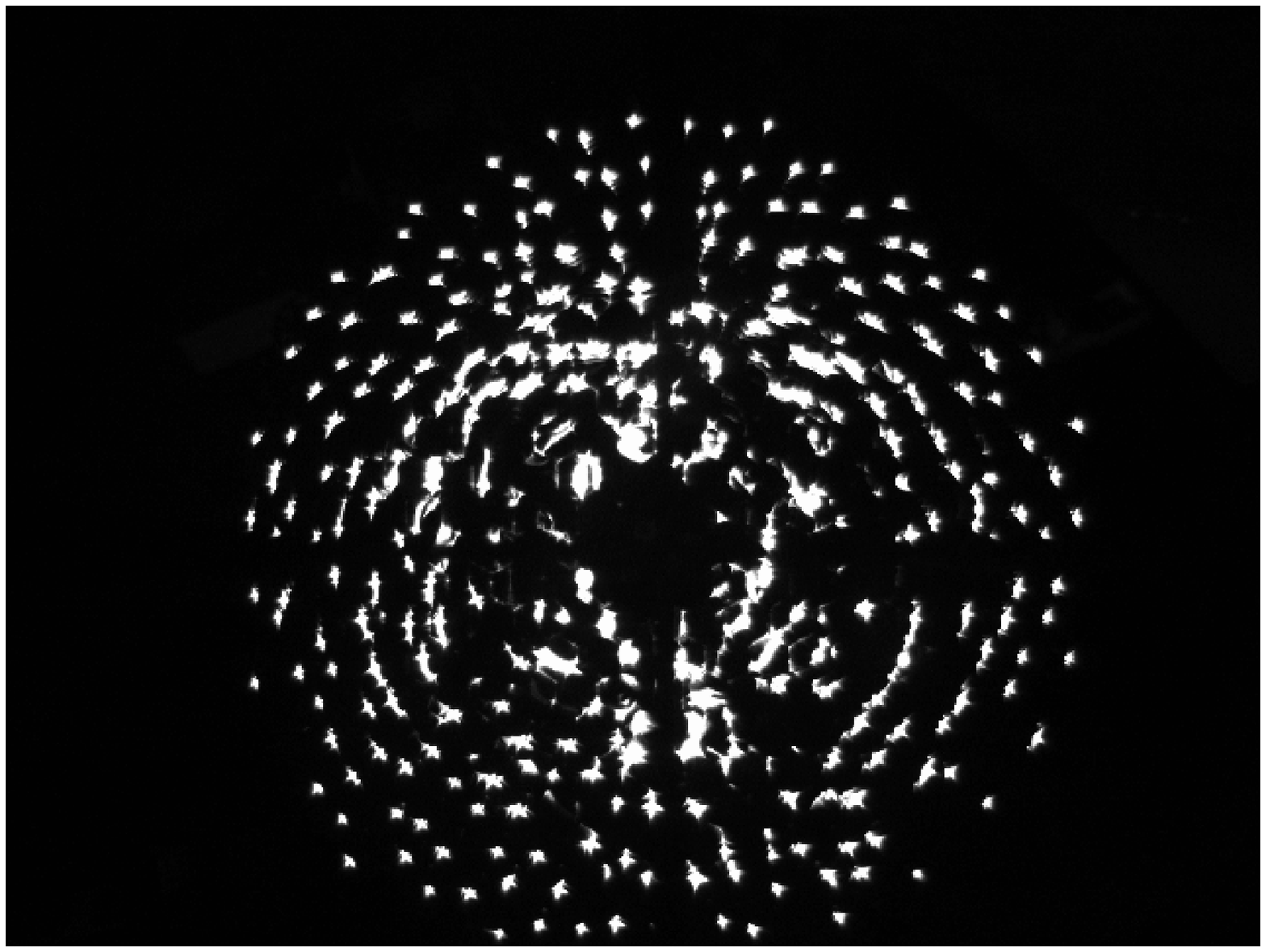}
}
\caption{ Panels (a) and (b) show the PSF of one of the VERITAS
  telescopes measured at $\sim$70$^{\circ}$ elevation, before and
  after implementation of the alignment corrections obtained using the
  system described in this article. The intensity values (greyscale)
  are plotted in arbitrary units.  The black circle indicates the size
  of a PMT in a VERITAS camera. PSF images are made by mounting a
  white screen on the telescope's focal plane and photographing, with
  a digital camera, the image of a bright star being tracked by the
  telescope.  The adjustments which were performed, and which led to
  the improved PSF, are plotted in Figure~\ref{fig:arrows}. Panels (c)
  and (d) show images of the reflector with the telescope pointing
  close to a star at $\sim$70$^{\circ}$ elevation, before and after
  implementation of alignment corrections.}
\label{fig:psfdish}
\end{figure}

\begin{figure}[]
\centering
\includegraphics[width=0.7\textwidth]{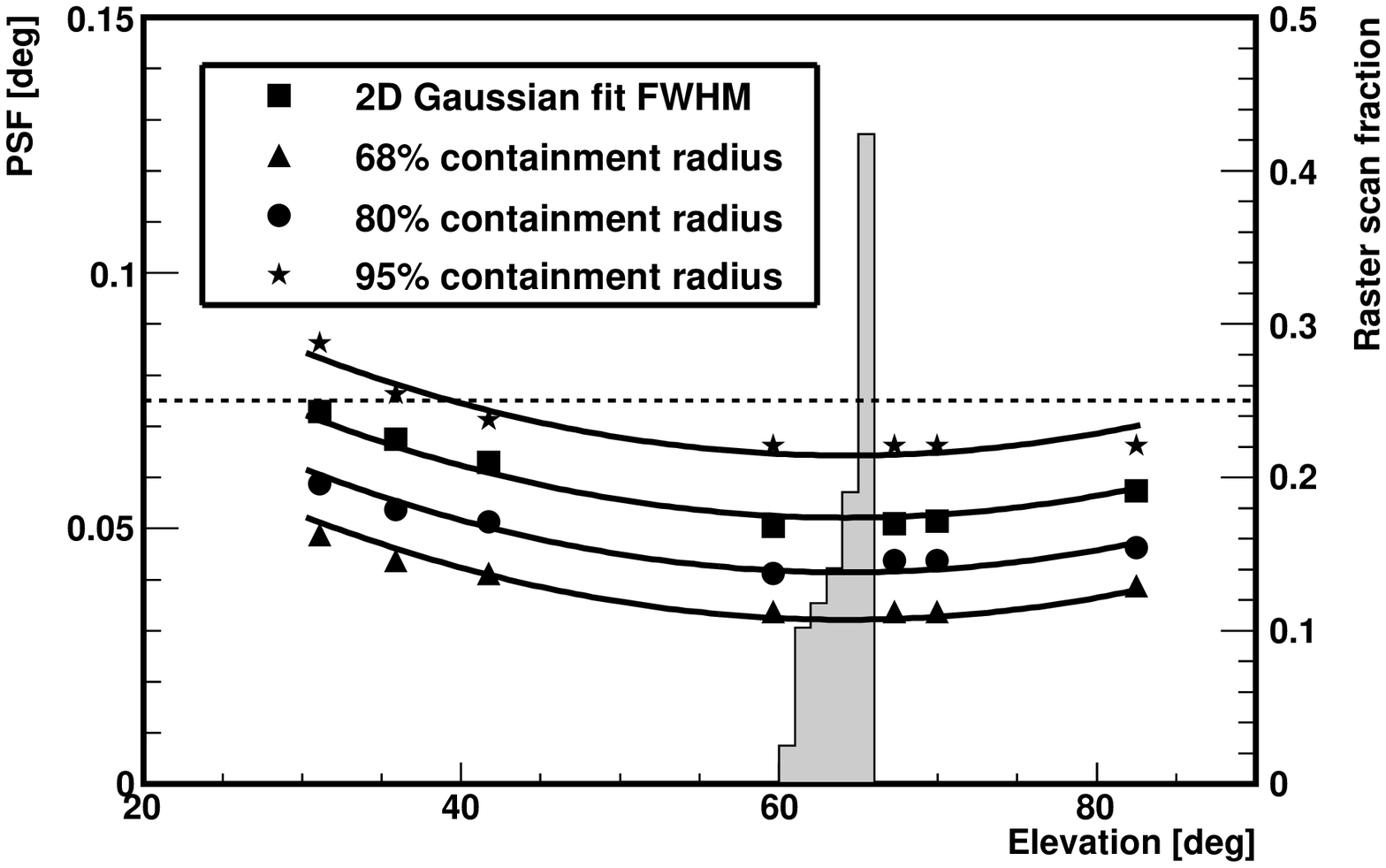}
\caption{A plot of the size of the point spread function against
  telescope elevation.  The horizontal dashed line represents the
  radius of the VERITAS PMTs.  The solid curves are the results of a
  quadratic fit to the data points, where the linear and quadratic
  coefficients were simultaneously fit to all four PSF measurement
  quantities.  The minimum of the fitted curves occurs at
  64.2$^{\circ}$.  The telescope was aligned using data acquired
  during a raster scan with mean elevation of 64.1$^{\circ}$.  The
  fraction of time the telescope spent tracking a given elevation
  during the raster scan is plotted in the grey histogram.  }
\label{fig:psfvel}
\end{figure}

Images of the PSF for a telescope, before and after the alignment, are
shown in Figure~\ref{fig:psfdish}. This figure also displays
corresponding images of the reflector, captured by the alignment
camera, while the telescope was tracking a bright star, before and
after the alignment corrections were applied.  Qualitative improvement
is evident.  The size of the PSF against elevation after the alignment
for one of the telescopes is plotted in Figure~\ref{fig:psfvel},
illustrating the elevation dependence of the PSF.  The elevation range
over which the raster scan was performed coincides with the elevation
of the smallest PSF size, as expected.

\section{Discussion}
The implementation of this alignment method has proven to be
successful.  The raster scan can be completed in two hours and the
data categorising the facet alignment have been shown to be both
useful and accurate.  The procedure is much easier to implement than
the previous alignment method \cite{Toner08} and is more
accurate. Moreover the mirror adjustments can be performed during
daylight hours alleviating competition for time working on the
telescope. Working during the day is also safer and easier for
observatory personnel.

During this initial implementation of the method we have not
investigated its limits.  We plan further tests in which we will
perform raster scans over a grid comparable in size to the now-reduced
PSF. We also hope to improve the accuracy with which we can measure the
mis-alignment angle by fitting the brightness distribution by a two
dimensional Gaussian function and using the fit centroid, rather than
the maximally brightened grid point, to identify the angle. To enable
us to implement finer facet adjustments anticipated from a high
resolution raster scan we have developed a âgearedâ
wrench with a ratio of four turns to one.  This will allow us to
reliably make adjustments as small as $\sim$0.0035$^{\circ}$
(corresponding to 1/32nd of a turn of an adjustment nut) to each
mirror.

The limiting value of the PSF depends on the positioning of the
mirrors, the positioning of the focal plane, the PSFs of the
individual facets and the spread in the size of the facet PSFs across
the mirror population. Ray-tracing simulations made assuming the
nominal telescope design specifications suggest that a PSF with an
80\% containment radius of 0.035$^{\circ}$ should be attainable.

Further to improving the PSF, we intend to use the alignment system to
better understand the flexure of the telescope OSS.  By performing
raster scans at several different elevations, the warping of the OSS
due to elevation changes can be measured.  These measurements may
point to possible modifications to stiffen the OSS and lessen the
elevation dependence of the PSF.

\section{Conclusion}
An alignment system based on the technique suggested by
\cite{Arqueros05} has been developed and used to improve the optics of
three VERITAS telescopes.  This has led to a reduction in the size of
the PSF by more than 30\%.  Moreover this system is less
labour-intensive than that which was previously used.  It has the
advantage that the telescope reflectors are directly optimised for use
at typical observing elevations.

Further investigations are planned.

\section{Acknowledgements}
VERITAS is supported by grants from the US Department of Energy, the
US National Science Foundation, and the Smithsonian Institution, by
NSERC in Canada, by Science Foundation Ireland, and by STFC in the UK.
We acknowledge the excellent work of the technical support staff at
the Fred Lawrence Whipple Observatory and the other institutions of
the VERITAS collaboration in the construction and operation of the
array.  In particular we would like to thank the personnel of the
Physics Department Machine Shop at McGill University for their part in
constructing the alignment tool.

We also gratefully acknowledge contributions from
V. Acciari, 
S. Fegan, 
K. Gibbs, 
G. Gillanders, 
R. Irvin, 
N. Karlsson, 
M. Lang,
J. Musser, 
J. Perkins,
A. Pichel, 
and 
S. Wissel.

\end{document}